\documentclass[a4paper,reqno,12pt,draft]{article}
\usepackage{amssymb,euscript}

\usepackage{epsfig}

\newcommand{\ie}{{\it i.e.}}

\newcommand{\be}{\begin{equation}}
\newcommand{\ee}{\end{equation}}
\newcommand{\bea}{\begin{eqnarray}}
\newcommand{\eea}{\end{eqnarray}}

\newcommand{\ft}{\footnote}

\begin{document}

\begin{flushright}
\end{flushright}
\begin{flushright}
\end{flushright}
\begin{center}

\LARGE{{\sc  The M5-brane on $K3\times T^2$}}\\
\bigskip
\Large{Neil Lambert}\ft{neil.lambert@kcl.ac.uk}\\
\bigskip
{Department of Mathematics\\
King's College\\
The Strand\\
London\\
WC2R 2LS, UK\\}


\end{center}


\bigskip
\begin{center}
{\bf {\sc Abstract}}
\end{center}

We discuss the low energy effective theory of an M5-brane wrapped on
a smooth holomorphic four-cycle of $K3\times T^2$, including the
special case of $T^6$. In particular we give the lowest order
equations of motion and resolve a puzzle concerning the counting of
massless modes that was reported in hep-th/9906094. In order to find
agreement with black hole entropy and anomaly inflow arguments we
propose that some of the moduli become massive.

\newpage

\section{Introduction}

One of the most exciting achievements in string theory is the
remarkable success in counting microscopic counting of black hole
states, starting with the work of \cite{Strominger:1996sh}. A
particularly elegant example of this is provided by considering an
M5-brane wrapped on a complex four-cycle of a Calabi-Yau
\cite{Maldacena:1997de}. This yields a black string in five
dimensions which can be further reduced to four dimensions by
wrapping the string on $S^1$ and including momentum along the $S^1$.
A notable feature of this analysis is that, for a generic
four-cycle, the M5-brane has a smooth worldvolume and hence the only
microscopic information needed is a knowledge of the worldvolume
fields and dynamics of a single M5-brane.

There have been several detailed accounts of the M5-brane wrapped on
cycles of a generic Calabi-Yau manifold, for example see
\cite{Vafa:1997gr,Minasian:1999qn,Lambert:1999ib,Minasian:1999gg,Bonelli},
and also the M5-brane on $K3$ \cite{Cherkis:1997bx}. The case that
we are interested in here concerns an M5-brane whose worldvolume has
non-trivial one-cycles which occurs when the Calabi-Yau degenerates
to $K3\times T^2$ or $T^6$ (see also \cite{Bonelli}). This situation
was discussed in \cite{LopesCardoso} where several puzzles arose. In
particular the number of massless states was not found to be in
accordance with $(0,4)$ supersymmetry and the counting of black hole
microstates failed (albeit at sub-leading order). To resolve these
problems the authors of \cite{LopesCardoso} proposed a novel
mechanism whereby some massless modes are charged with respect to
the worldvolume gauge fields that arise from reduction of the
two-form. The main purpose of this paper is to investigate this
proposal. However we find that the correct resolution comes from
including additional massless modes which are present when the
Calabi-Yau is $K3\times T^2$ or $T^6$.

The rest of this paper is organized as follows. In section two we
present the lowest order equations of motion for an M5-brane which
is wrapped on a smooth cycle $P$ in spacetime. In section three we
consider in detail the case where spacetime is of the form ${\cal M}
= {\bf {R}}^{1,4}\times K3\times T^2$ and ${\cal M} = {\bf
{R}}^{1,4}\times T^6$. We provide a careful counting of the normal
bundle moduli and resolve a puzzle concerning $(0,4)$ supersymmetry
that was observed in \cite{LopesCardoso}. In section four we
consider four-dimensional black hole states that arise by further
compactification on $S^1$. We find that, using our analysis, the
usual counting of left-moving massless modes to determine black hole
entropy does not agree with the supergravity calculations or
arguments using anomalies. To resolve this discrepancy we propose
that  $h_{1,0}(P)$ $(4,4)$ multiplets must become massive and hence
do not appear in the low energy effective action. This provides an
alternative resolution to a second puzzle discussed in
\cite{LopesCardoso}. Finally section five contains a brief
conclusion.

\section{Lowest Order Equations of Motion}

Covariant  equations of motion of the M5-brane were first derived in
\cite{Howe}. We will not need give the full non-linear form of these
equations, however it will be enlightening to give the lowest order
equations (in terms of a derivative expansion). We will work in
static gauge where the six coordinates $x^\mu$,
$\mu,\nu=0,1,2,...,5$, of the M5-brane worldvolume are identified
with the first six coordinates of spacetime. The massless fields
consist of 5 scalars $X^A$, $A,B=6,7,8,...,10$, a two-form
$B_{\mu\nu}$ and a Fermion $\psi$ which satisfies
$\Gamma_{012345}\psi=-\psi$. Here we use a full 32-component spinor
of $SO(1,10)$. It will be sufficient to work at the lowest order in
the fields. $X^A$ represents the coordinates of the M5-brane in the
transverse space and in particular $X^A=0$ corresponds to the
M5-brane wrapped on a calibrated submanifold. We use
$M,N=0,1,2,...,10$ to denote all eleven coordinates. We use an
underline to denote tangent space indices. We will use a hat to
denote eleven-dimensional quantities, the spacetime is denoted by
${\cal M}$ and the M5-brane worldvolume by $\cal W$.

First recall the case where the worldvolume $\cal W$ admits a chiral
Killing spinor $\epsilon$; $D_\mu \epsilon=0$,
$\Gamma_{012345}\epsilon=\epsilon$. For example if ${\cal M} = {\bf
R}^{1,4}\times K3\times T^2$ and ${\cal W}={\bf R}^{1,1}\times K3$.
To lowest order in fluctuations, the equations of motion are just
that of a free theory on a curved background
\begin{eqnarray}\label{oldeqs}
\nonumber
D^2 X^{\underline A} &=&0\\
i\Gamma^\mu D_\mu\psi &=&0\\
\nonumber
H_{\mu\nu\lambda}&=&\frac{1}{3!}\epsilon_{\mu\nu\lambda\rho\sigma\tau}H^{\rho\sigma\tau}
,
\end{eqnarray}
where $H_{\mu\nu\lambda} = 3\partial_{[\mu}B_{\nu\lambda]}$ and
$\epsilon_{\mu\nu\lambda\rho\sigma\tau} $ is totally antisymmetric
with  $\epsilon^{\underline{012345}}=1$. These equations are
invariant under the supersymmetry transformations
\begin{eqnarray}
\nonumber \delta X^{\underline A} &=& i\bar\epsilon\Gamma^{\underline A}\psi\\
\delta B_{\mu\nu} &=& i\bar\epsilon\Gamma_{\mu\nu}\psi\\
\nonumber \delta\psi &=& \partial_\mu X^{\underline
A}\Gamma^\mu\Gamma_{\underline A}\epsilon + \frac{1}{2\cdot
3!}\Gamma^{\mu\nu\lambda}H_{\mu\nu\lambda}\epsilon .
\end{eqnarray}

Next we consider the case where  the spacetime ${\cal M}$ admits a
chiral covariantly constant spinor $\hat \epsilon$, $\hat D_M\hat
\epsilon=0$, $\Gamma_{012345}\hat\epsilon=\hat\epsilon$ but where
this does not descend to a Killing spinor on $\cal W$. For example
we can take ${{\cal M}} ={\bf R}^{1,4}\times K3\times T^2$ but with
${\cal W} = {\bf R}^{1,1}\times \Sigma\times T^2$ where $\Sigma$ is
a 2-cycle in $K3$. We choose a vielbein frame such that, at least
locally,
\begin{equation}
\hat e_M^{\ \underline N} = \left(\matrix{ e_\mu^{\ \ \underline
\nu} &0\cr e_A^{\ \ \underline\nu}&e_A^{\ \ \underline
B}\cr}\right).
\end{equation}
Therefore,  in the static gauge that we are considering, the induced
metric on the M5-brane is simply $g_{\mu\nu} = \hat
g_{\mu\nu}(X^{\underline A}=0)$. We may further choose $\hat
\omega_\mu^{\ \underline {\nu B}}(X^{\underline A}=0)=0$ and $\hat
\omega_\mu^{\ \underline{\nu\lambda}}(X^{\underline A}=0) =
\omega_\mu^{\ \underline{\nu\lambda}}$, where $\omega_\mu^{\
\underline{\nu\lambda}}$ is the spin connection that one would
calculate from the vielbein $e_\mu^{\ \underline \nu}$. Finally we
also see that  $\hat \Gamma_\mu = \hat e_\mu^{\ \underline
\nu}\Gamma_{\nu} = \Gamma_\mu$ is the same $\gamma$-matrix that one
would calculate simply using the worldvolume metric $g_{\mu\nu}$.

This allows us reinterpret the bulk Killing spinor condition on the
worldvolume as
\begin{eqnarray}
\nonumber 0&=&\hat D_\mu \epsilon \\
&=& \partial_\mu\epsilon + \frac{1}{4}\hat\omega_\mu^{\
\underline{\nu\lambda}}\Gamma_{\underline{\nu\lambda}}+
\frac{1}{4}\hat\omega_\mu^{\
\underline{AB}}\Gamma_{\underline{AB}}\\
\nonumber &=& D_\mu\epsilon + A_\mu\epsilon,
\end{eqnarray}
where $\epsilon = \hat \epsilon (X^{\underline A}=0)$,
$\omega_{\mu}^{\ \underline{AB}}=\hat \omega_{\mu}^{\
\underline{AB}}(X^{\underline A}=0)$ and $A_\mu =
\frac{1}{4}\omega_{\mu}^{\ \underline{AB}}\Gamma_{\underline{AB}}$.

We find that, at lowest order in the fields $X^A$, $B_{\mu\nu}$ and
$\psi$, the following symmetries close on-shell into translations,
gauge transformations and local tangent frame rotations
\begin{eqnarray}\label{newsusy}
\nonumber \delta X^{\underline A} &=& i\bar\epsilon\Gamma^{\underline A}\psi\\
\delta B_{\mu\nu} &=& i\bar\epsilon\Gamma_{\mu\nu}\psi\\
\nonumber \delta\psi &=& \nabla_\mu X^{\underline
A}\Gamma^\mu\Gamma_{\underline A}\epsilon + \frac{1}{2\cdot
3!}\Gamma^{\mu\nu\lambda}H_{\mu\nu\lambda}\epsilon,
\end{eqnarray}
where $\nabla_\mu X^{\underline A} = \partial_\mu X^{\underline A} +
\omega_{\mu \underline B}^{\ \ \ \underline A}X^{\underline B}$. The
Fermion equation of motion that is required to close the algebra is
\begin{equation}\label{Feq}
\Gamma^\mu \nabla_\mu\psi=0,
\end{equation}
where $\nabla_\mu\psi = D_\mu\psi + A_\mu\psi$.

What are the remaining equations of motion? The $B$-field has a
self-dual field strength $H= dB$ and hence one finds $d\star H=0$.
This condition is preserved by the supersymmetries (\ref{newsusy}).
Taking a supersymmetry variation of the Fermion equation of motion
(\ref{Feq}) leads the condition
\begin{equation}
\label{Fvar} 0 = \Gamma_{\underline A} \nabla^2 X^{\underline
A}\epsilon +\frac{1}{2}\Gamma_{\underline
A}\Gamma^{\mu\nu}F_{\mu\nu\underline B}^{\ \ \ \ \underline A}
X^{\underline B}\epsilon,
\end{equation}
where
\begin{eqnarray}
\nonumber F_{\mu\nu\underline B}^{\ \ \ \ \underline
A}&=&\partial_{\mu}\omega_{\nu\underline B}^{\ \ \ \underline A}-
\partial_{\nu}\omega_{\mu\underline B}^{\ \ \ \underline A}
+\omega_{\mu\underline B}^{\ \ \ \underline C}\omega_{\mu\underline
C}^{\ \ \ \underline A}-\omega_{\mu\underline B}^{\ \ \ \underline
C}\omega_{\mu\underline C}^{\ \ \ \underline A}\\
&=&\hat R_{\mu\nu\underline B}^{\ \ \ \ \ \underline
A}(X^{\underline A}=0).
\end{eqnarray}
To proceed we assume there is a relation of the form
\begin{equation}\label{Mdef}
\frac{1}{2}\Gamma_{\underline A}\Gamma^{\mu\nu}F_{\mu\nu\underline
B}^{\ \ \ \ \underline A}\epsilon = M^{\underline A}_{\ \underline
B}\Gamma_{\underline A}\epsilon,
\end{equation}
in which case the equation of motion for $X^{\underline A}$, along
with the other fields, is
\begin{eqnarray}\label{Xeq}
\nonumber \nabla^2X^{\underline A} + M^{\underline A}_{\ \underline
{B}}X^{\underline B}&=&0\\
i\Gamma^\mu \nabla_\mu\psi &=&0\\
\nonumber
H_{\mu\nu\lambda}&=&\frac{1}{3!}\epsilon_{\mu\nu\lambda\rho\sigma\tau}H^{\rho\sigma\tau}.
\end{eqnarray}

We need to confirm that these equations are supersymmetric. To this
end we note that the Fermion equation of motion (\ref{Feq}) implies
\begin{equation}
\nabla^2 \psi -\frac{1}{4} R\psi
+\frac{1}{8}\Gamma^{\mu\nu}F_{\mu\nu}^{\ \
\underline{CD}}\Gamma_{\underline{CD}}\psi=0.
\end{equation}
To connect with (\ref{Fvar}) we multiply this on the left by
$\bar\epsilon\Gamma^{\underline A}$ to find
\begin{equation}\label{Feqcon}
\bar\epsilon\Gamma^{\underline A}\nabla^2 \psi -\frac{1}{4}
R\bar\epsilon\Gamma^{\underline A}\psi
+\frac{1}{8}\bar\epsilon\Gamma^{\underline
A}\Gamma^{\mu\nu}F_{\mu\nu}^{\ \
\underline{CD}}\Gamma_{\underline{CD}}\psi=0.
\end{equation}
Next we note that since $\hat \omega_\mu^{\ \underline{\nu A}}=0$ we
have that $\hat R_{\mu\nu}^{\ \ \underline{\lambda A}}=0$ and
therefore the Killing spinor integrability condition $[\hat
D_\mu,\hat D_\nu]\epsilon = \frac{1}{4}\hat R_{\mu\nu}^{\ \
\underline{MN}}\Gamma_{\underline{MN}}\epsilon=0$ implies
\begin{equation}
 R_{\mu\nu}^{\ \
 \underline{\lambda\rho}}\Gamma_{\underline{\lambda\rho}}\epsilon =
 -F_{\mu\nu}^{\ \
 \underline{CD}}\Gamma_{\underline{CD}}\epsilon,
\end{equation}
and hence
\begin{equation}\label{newblah}
\bar\epsilon R
=\frac{1}{2}\bar\epsilon\Gamma_{\underline{CD}}\Gamma^{\mu\nu}F_{\mu\nu}^{\
\ \underline{CD}}.
\end{equation}
Using this we see that (\ref{Feqcon}) implies
\begin{equation}\label{blah}
\bar\epsilon \Gamma^{\underline A}\nabla^2 \psi
+\frac{1}{2}\bar\epsilon\Gamma^{\underline
B}\Gamma^{\mu\nu}F_{\mu\nu\underline B}^{\ \ \ \ \underline A}
X^{\underline B}\psi=0.
\end{equation}
The $X^{\underline A}$ equation can be compared to (\ref{blah}) by
noting that
\begin{equation}
\delta\nabla^2 X^{\underline A} = i\bar\epsilon\Gamma^{\underline
A}\nabla^2\psi,
\end{equation}
and one sees that the equations (\ref{Xeq}) are preserved by
supersymmetry.

Let us consider  for example the case where $\cal W$ is
non-trivially embedded in eight dimensions, so that only
$F_{\mu\nu}^{\ \ \underline{67}}\ne 0$. We see from (\ref{Mdef}) and
(\ref{newblah}) that the only non-vanishing components of
$M^{\underline A}_{\ \underline B}$ are
\begin{equation}
M^{\underline 6}_{\ \underline 6}= M^{\underline 7}_{\ \underline
7}= R.
\end{equation}
Thus we find the scalar equations are
\begin{eqnarray}\label{MXeqs}
\nonumber \nabla^2
X^{\underline 6}+RX^{\underline 6}&=& 0\ , \\
\nabla^2 X^{\underline 7}+RX^{\underline 7}&=&  0\ ,\\
\nonumber \nabla^2 X^{\underline A}&=& 0\ , \qquad A=8,9,10.
\end{eqnarray}

\section{Counting Moduli}

In the previous section we determined the lowest order equation of
motion for an M5-brane wrapped on a general calibrated submanifold
$\cal W$ of $\cal M$. As a result we saw that the Fermions and
scalar fields couple minimally to the gauge field associated to the
structure group of the normal bundle and some scalars develop a mass
term from the curvature. However the three-form remains closed and
self-dual (at the linearized level). In this section we wish to
perform a precise counting of the massless degrees of freedom  for
an M5-brane wrapped on a four-cycle $P\subset {\cal M}$, \ie\ ${\cal
W} = {\bf R}^{1,1}\times P$, in a spacetime of the form ${\cal M} =
{\bf R}^{1,4}\times {\cal K}$ where $\cal K$ is some compact
Calabi-Yau space that contains $P$. This has been discussed in great
detail in \cite{Maldacena:1997de,LopesCardoso} and we will largely
follow their discussion.

The simplest field to consider is the dimensional reduction of the
two-form gauge field. As a consequence of the self-duality condition
one finds $b_2^+(P)$ right moving scalars and $b_2^-(P)$ left moving
scalars. For the compact K\"ahler manifolds that we consider here
$b_2^+(P) =2h_{2,0}(P)+1 $ and $b_2^-(P)=h_{1,1}(P)-1$. If
$h_{1,0}(P)$ is non-vanishing then there will be $2h_{1,0}(P)$
Abelian gauge fields in the two-dimensional effective theory.
However these are non-dynamical we will not need them here.

Next we consider reduction of the scalars $X^{\underline A}$. In
total there are five. Three of these, $X^8,X^9,X^{10}$ simply
parameterize the location in the non-compact transverse space. These
always give 3 left and 3 right moving scalars in two dimensions. The
remaining two scalars are in fact sections of the normal bundle of
$P$ inside $\cal K$. As such the number of such zero modes is hard
to calculate. Let us denote the number of normal bundle moduli by
$N(P,{\cal K})$. These are left-right symmetric and we will discuss
them in more detail shortly.

As for the Fermions it is well known (see \cite{Green:1987mn}) that
spinors on a K\"ahler manifold $P$ can be realized as $(0,p)$-forms
on $P$. To see this one first consider complex coordinates for $P$
so that $\{\Gamma^a,\Gamma^b\}=\{\Gamma^{\bar a},\Gamma^{\bar
b}\}=0$ and $\{\Gamma^a,\Gamma^{\bar b}\}=2g^{a\bar b}$ with
$a,b=z,w$. In particular we consider a spinor ground state
$|0\rangle$ which is annihilated by the holomorphic
$\gamma$-matrices; $\Gamma^a|0\rangle=0$. We can then construct a
general spinor by
\begin{equation}
|\psi> = \omega |0\rangle + \Gamma^{\bar a}\omega_{\bar
a}|0\rangle+\frac{1}{2}\Gamma^{\bar a\bar b}\omega_{\bar a\bar
b}|0\rangle.
\end{equation}
By construction $\omega_{\bar a_1...\bar a_p}$ is totally
anti-symmetric and hence represents a $(0,p)$-form on $P$.
Furthermore if we choose the complex $\gamma$-matrices
$\Gamma^z=\Gamma^2+i\Gamma^4$ and $\Gamma^w=\Gamma^3+i\Gamma^5$ then
one sees that $\Gamma_{2345}|0\rangle =|0\rangle$ and more generally
\begin{equation}
\Gamma_{2345}|\omega_p\rangle = (-1)^p|\omega_p\rangle,
\end{equation}
where $|\omega_p\rangle = \frac{1}{p!}\omega_{\bar a_1...\bar
a_p}\Gamma^{\bar a_1..\bar a_p}|0\rangle$. Since  the Fermions on
the M5-brane satisfy $\Gamma_{012345}\psi=-\psi$ we see that
$|\omega_p\rangle$ leads to right and left moving Fermions in two
dimensions if $p$ is even or odd respectively.

To find massless two-dimensional modes we assume that $|0\rangle$ is
Killing with respect to $\nabla$ defined above. In this case one see
that solutions to $(\Gamma^a\nabla_a+\Gamma^{\bar a}\nabla_{\bar
a})\psi=0$ correspond to $\bar \partial_{[\bar b} \omega_{\bar
a_1...\bar a_p]} =0$ and $g^{b\bar a_1}\partial_{[b}\omega_{\bar
a_1...\bar a_p]} =0$, \ie\ $\omega_p \in H^{(0,p)}(P)$. Thus one
finds that number of massless left and right moving two-dimensional
Fermions is
\begin{equation}
N_F^L = 4h_{1,0}(P), \qquad N_F^R = 4(h_{0,0}(P)+h_{2,0}(P)).
\end{equation}
Here the factor of 4 comes from the fact  the spinor `groundstate'
$|0\rangle$ can be thought of as having 32 real components but is
subject to the three constraints:
$\Gamma^z|0\rangle=\Gamma^w|0\rangle=0$ and
$\Gamma_{012345}|0\rangle=-|0\rangle$. Thus $|0\rangle$ has four
real independent components.

Let us summarize our counting so far. We find
\begin{eqnarray}\label{Ncount}
\nonumber {N}_B^L&=&  2+  h_{1,1}(P)+N(P,{\cal K})\\
\nonumber {N}_B^R &=& 4+2h_{2,0}(P)+  N(P,{\cal K}) \\
  {N}_F^L &=& 4h_{1,0}(P) \\
  \nonumber {N}_F^R &=& 4h_{2,0}(P)+4,
\end{eqnarray}
where we have assumed that $h_{0,0}(P)=1$. Since the wrapped
M5-brane preserves (at least) $(0,4)$ supersymmetry the right-movers
must have Bose-Fermi degeneracy. This immediately allows us to
determine the number of normal moduli to be
\begin{equation}\label{newnormal}
N(P,{\cal K}) = 2h_{2,0}(P),
\end{equation}
and hence the massless spectrum is
\begin{eqnarray}\label{ourcount}
\nonumber N_B^L&=&  2h_{2,0}(P)+h_{1,1}(P)+2\\
\nonumber N_B^R &=&   4h_{2,0}(P)+4\\
  N_F^L &=&  4h_{1,0}(P)\\
  \nonumber N_F^R &=& 4h_{2,0}(P)+4.
\end{eqnarray}
Note that this also ensures that the number of right-moving modes is
a multiple of 4, as also required by $(0,4)$ supersymmetry. We would
like to emphasis that this formula should apply whenever it make
sense to talk of a classical M-brane that is wrapped on a smooth
complex submanifold of any smooth Calabi-Yau (including $K3\times
T^2$ and $T^6$).

This formula should be contrasted with the result
\begin{equation}\label{normal}
N(P,{\cal K}) = 2h_{2,0}(P) - 2h_{1,0}(P),
\end{equation}
first obtained in \cite{Maldacena:1997de} for an ample four-cycle
$P$ in a  generic Calabi-Yau and extended to $K3\times T^3$ and
$T^6$ in \cite{LopesCardoso}. We see that there is agreement for a
generic Calabi-Yau where $h_{1,0}(P)=0$. However, as pointed out in
\cite{LopesCardoso}, the formula (\ref{normal}) contradicts
supersymmetry when $h_{1,0}(P)\ne 0$. In the rest of this section we
will argue that (\ref{newnormal}) is the correct counting and
identify the missing modes that are absent from (\ref{normal}).

We start with a brief review of  the calculation in
\cite{Maldacena:1997de}. This starts from the observation that a
4-cycle $P\subset {\cal K}$ is defined by the zeros of a section of
a line bundle over ${\cal K}$. The Poincar\'e dual two-form to $P$,
which we denote by $[P]$, determines the Chern class of the line
bundle. Thus counting the number of deformations of $P$ corresponds
to counting the (real) dimension of the dimension of the space of
line bundles. However one must take into account the fact that if
$P$ is described by zeros of a section $s$ then the zeros of
$\lambda s$ describe the same $P$ for any $\lambda \in {\bf
C}^\star$. Thus one needs the real dimension of the projective space
of line bundles. In this way one determines $N(P,{\cal K})$ through
\begin{eqnarray}\label{indexcal}
\nonumber N(P,{\cal K}) &=& 2{\rm dim}(H^0({\cal K},{\cal L}))-2\\
\nonumber &=&2\sum _i (-1)^i{\rm dim}(H^i(P,{\cal L}))-2\\
&=& 2\int_{\cal K} e^{[P]}{\rm Td}({\cal K})-2\\
\nonumber &=& \frac{1}{3}\int_{\cal K} [P]^3+\frac{1}{6}\int_{\cal
K}[P]\wedge c_2({\cal K})-2.
\end{eqnarray}
Here the second line follows from the Kodaira vanishing theorem;
$H^i({\cal K},{\cal L})=\emptyset$ for $i>0$, and the third line
from a Riemann-Roch index formula. For an account of these theorems
see \cite{Hirzebruch,GH}. Next one can use the formula (see
\cite{Maldacena:1997de,LopesCardoso})
\begin{equation}\label{h20}
h_{2,0}(P) = \frac{1}{6}\int_{\cal K} [P]^3 + \frac{1}{12}\int_{\cal
K} [P]\wedge c_{2}({\cal K}) + h_{1,0}(P)-1,
\end{equation}
to obtain (\ref{normal}).

So what is missing from this calculation (\ref{indexcal})? A central
assumption of \cite{Maldacena:1997de} is that $P$ is an ample cycle.
Technically this means that the Poincar\'e dual two-form $[P]$ lies
inside the K\"ahler cone, \ie\ it defines a positive volume for all
complex 2-,4- and 6-cycles in $\cal K$. More intuitively an ample
cycle $P$ of a manifold $\cal K$ is one that is sufficiently generic
so that the set of all normal vectors to $P$ spans the entire
tangent space of $\cal K$.

A key assumption of the Kodaira vanishing theorem is that the line
bundle $\cal L$ is positive and hence (\ref{indexcal}) counts the
dimension of the space of positive line bundles. While every ample
four-cycle in $\cal K$ defines a positive line bundle there are zero
modes which do not correspond to positive line bundles. In
particular consider translations of $P$ along any of the $S^1$
factors in ${\cal K}$. These $S^1$ factors are trivial and
describing the location of an M5-brane in $S^1$ simply corresponds
to specifying a value of the coordinate for that $S^1$. As such the
location is simply a section of a trivial $U(1)$ line bundle over
$P$ and this extends to a trivial $U(1)$ bundle over $\cal K$. These
deformations  are not counted in (\ref{normal}) since the associated
line bundle is trivial. There are $2h_{1,0}({\cal K})$ such
translations and, using the Lefschetz hyperplane theorem (valid for
ample four-cycles), we have that $h_{1,0}({\cal K})=h_{1,0}(P) $.
Therefore we find an extra $2h_{1,0}(P)$ normal modes that arise
from translations along the $S^1$ factors of $\cal K$. Including
these modes in (\ref{indexcal}) gives (\ref{newnormal}).

An alternative description of these translational modes is to note
that the $S^1$ factors are orbits of a $U(1)$ Killing isometry that
acts on ${\cal K}$. An ample cycle breaks the symmetries
corresponding to translations along the $S^1$ factors and hence
there must be $2h_{1,0}({\cal K}) = 2h_{1,0}(P)$ Goldstone modes.
There are also smooth but non-ample four-cycles for which $
h_{1,0}(P) \ne h_{1,0}({\cal K})$ and the index theorem does not
apply. In these cases one also finds that the cycle breaks fewer
$U(1)$ isometries and as a result has fewer Goldstone modes. We will
explicitly see in the examples below that nevertheless
(\ref{newnormal}) is valid for all smooth four-cycles, as required
by supersymmetry.

\subsection{Three Examples}

To illustrate this discussion let us  consider some explicit
examples for ${\cal K}= K3\times T^2$. We will consider three
choices for $P$: $P=K3$, $P=\Sigma\times T^2$ and $P=K3+
\Sigma\times T^2$, where $\Sigma$ is a two-cycle in $K3$. For a
useful account of various facts about $K3$ see
\cite{Aspinwall:1996mn}. Following this we will also discuss the
case where ${\cal K}= T^6$.

First we consider the case where $P=K3$ which was first studied in
detail in \cite{Cherkis:1997bx}. The Hodge diamond of $K3$ is
\begin{equation}
K3:\qquad\matrix{&&1&&\cr &0&&0&\cr 1&&20&&1\cr&0&&0\cr &&1&&\cr}.
\end{equation}
In this case it is clear that $N(K3,K3\times T^2)=2$ since $\cal K$
is simply a direct product ${\cal K} = K3\times T^2$. Hence the
normal bundle to $P=K3$ is trivial and there is no obstruction to
moving the $K3$ around inside $\cal K$. Since the Killing spinor on
$K3$ is chiral, reduction on $K3\times T^2$ leads to a
two-dimensional theory with $(0,8)$ supersymmetry. Looking at the
field content we find the massless modes
\begin{equation}
N_B^L = 24, \qquad N_B^R = 8\ ,\qquad N_F^L = 0, \qquad N_F^R = 8,
\end{equation}
which is the same as the worldsheet action for the Heterotic string
on $T^3$.

Next we consider the case where $P=\Sigma\times T^2$. Let us suppose
that $\Sigma$ is ample in $K3$. In complex dimension two the
Lefschetz hyperplane theorem does not imply that
$h_{1,0}(\Sigma)=h_{1,0}(K3)=0$ and hence $h_{1,0}(\Sigma)=g$ need
not be zero.  Assuming $\Sigma$ is connected the Hodge diamond of
$\Sigma\times T^2$ is
\begin{equation}\label{Hodge}
\Sigma\times T^2:\qquad\matrix{&&1&&\cr &1+g&&1+g&\cr
g&&2+2g&&g\cr&1+g&&1+g\cr &&1&&\cr}.
\end{equation}
To determine $N(\Sigma\times T^2,K3\times T^2)$ we note that
$N(\Sigma\times T^2,K3\times T^2) = N(\Sigma,K3)$. Since $K3$ does
not have any $S^1$ factors we may use a similar calculation as in
(\ref{indexcal}), suitably adapted to 2 complex dimensions. We find
that
\begin{eqnarray}\label{K3calc}
\nonumber {\rm dim}(H^0(K3,{\cal L}))&=& \sum_{i=0}^2 (-1)^i {\rm dim}(H^i(K3,{\cal L})) \\
&=& \int_{K3} e^{[\Sigma]}{\rm Td}(K3)\\
\nonumber &=& \int_{K3}\frac{1}{12}c_2(K3) + \frac{1}{2}[\Sigma]^2\\
\nonumber &=& 2 + \frac{1}{2}\int_{K3}[\Sigma]^2.
\end{eqnarray}
Thus proceeding as before and taking account of the projective
equivalence we find
\begin{equation}
N(\Sigma,K3) = 2 ({\rm dim}(H^0(K3,{\cal L}))-1) = 2
+\int_{K3}[\Sigma]^2.
\end{equation}
Continuing we observe that $[\Sigma]$ is the Poincare dual to
$\Sigma$ and hence we find
 \begin{eqnarray}\label{NK3}
 \nonumber \int_{K3}[\Sigma]^2  &=& \int_{\Sigma}[\Sigma]\\
&=& -\int_{\Sigma} c_1({\cal L})\\
\nonumber &=& 2g-2,
\end{eqnarray}
where in the second line we have used the adjunction formula to
identify $[\Sigma]=-c_1({\cal L})$ and the last line follows from
the well known formula for the Euler number of a two-dimensional
surface. Note that ample implies that $g\ge 2$. Thus
\begin{equation}
N(\Sigma\times T^2,K3\times T^2)=N(\Sigma,K3) = 2g.
\end{equation}
Putting this all together we see that the field content is
\begin{equation}
N_B^L = 4+4g\qquad N_B^R =4+4g\qquad N_F^L =4+4g\qquad N_F^R = 4+4g.
\end{equation}
Note that the spectrum is non-chiral which is a consequence of the
fact that in this case $(4,4)$ supersymmetry is preserved.

The previous two cases are not generic and in particular the cycle
$P$ is not ample in $\cal K$. In these cases the formula
(\ref{normal}) does not necessarily apply and indeed it doesn't
always agree with our results. However the formula (\ref{newnormal})
is valid and agrees with our discussion. Our final case $P=K3+
\Sigma\times T^2$ is generic in that $P$ is ample. Therefore the
calculation (\ref{indexcal}) is valid. However since $h_{1,0}(P)\ne
0$ we will be able to test whether (\ref{normal}) or
(\ref{newnormal}) reproduces the correct number of normal modes. To
see what these formulae give us we need to compute $h_{2,0}(P)$.
From (\ref{h20}) we find
\begin{equation}
h_{2,0}(P) = \frac{1}{6}\int_{K3\times T^2} [P]^3 +
\frac{1}{12}\int_{K3\times T^2} [P]\wedge c_{2}(K3\times T^2) +
h_{1,0}(P)-1.
\end{equation}
Writing $[P]=dvol_{T^2}+[\Sigma]$, where $dvol_{T^2}$ is the unit
volume form of $T^2$, we find
\begin{eqnarray}\label{ghgh}
\nonumber h_{2,0}(P) &=& \frac{1}{2}\int_{K3}[\Sigma]^2 + \frac{1}{12}\int_{K3}c_2(K3)+h_{1,0}(P)-1 \\
  &=& g+h_{1,0}(P),
\end{eqnarray}
where we have used (\ref{NK3}) and $\chi(K3)=24$ in the second line.
Thus since $P$ is ample $h_{1,0}(P)=1$ and (\ref{normal}) predicts
$2g+2$ normal modes and (\ref{normal}) only $2g$ normal modes.
However one expects that there are always two normal modes which
come from translations along $T^2$ and also that all of the normal
modes which exist in the embedding of $\Sigma$ in $K3$ should also
exist here. This example therefore demonstrates that the formula
(\ref{normal}) fails to include the translational modes whereas
(\ref{newnormal}) correctly accounts for all zero-modes.

For completeness we note that \cite{Maldacena:1997de,LopesCardoso}
\begin{eqnarray}\label{gghh}
\nonumber h_{1,1}(P) &=& \frac{2}{3}\int_{K3\times T^2} [P]^3 + \frac{5}{6}\int_{K3\times T^2} [P]\wedge c_{2}(K3\times T^2)+2h_{1,0}(P)\\
  &=& 4g+16+2h_{1,0}(P).
\end{eqnarray}
Thus  we find, from (\ref{ourcount}),
\begin{equation}
N_B^L = 6g+22, \qquad N_B^R =4g+8 \ ,\qquad N_F^L = 4, \qquad N_F^R
= 4g+8.
\end{equation}

Finally we can follow the above discussion and consider what happens
to these three cases when $K3$ is replaced by $T^4$, \ie\ ${\cal
K}=T^6$. In the first case where $P=T^4$ the same arguments give
$N(T^4,T^6)=2$ and hence
\begin{equation}
N_B^L = 8, \qquad N_B^R =8 \ ,\qquad N_F^L = 8, \qquad N_F^R = 8,
\end{equation}
which of course is just a straightforward reduction of the M5-brane
on $T^4$ and has $(8,8)$ supersymmetry.

In the second case all we need to do is replace $c_2(K3)=24$ by
$c_2(T^4)=0$ in (\ref{K3calc}) and we now find that the index
theorem gives $2{\rm dim}(H^0(\Sigma,T^4)-2=2g-4$ normal modes.
However we claim that, in addition to the modes counted by the index
theorem, to obtain the number of normal mode deformations of
$\Sigma$ inside $T^4$ we must also include  4 translational
Goldstone modes and hence $N(\Sigma,T^4)=2g$. Thus we find
$N(\Sigma\times T^2,T^6)=N(\Sigma,T^4)=2g$. From (\ref{Hodge}) we
find the total spectrum is
\begin{equation}
N_B^L = 4g+4, \qquad N_B^R =4g+4 \ ,\qquad N_F^L = 4g+4, \qquad
N_F^R = 4g+4.
\end{equation}
Just as in the $K3\times T^2$ case this spectrum is non-chiral as
result of enhanced $(4,4)$ supersymmetry.

In the third case where $P = T^4 + \Sigma\times T^2$ the cycle is
ample we have that $h_{1,0}(P)=3$. The calculations (\ref{ghgh}) and
(\ref{gghh}) give (replacing $c_2(K3)=24$ by $c_2(T^4)=0$ and
setting $h_{1,0}(P)=3$)
\begin{equation}
h_{2,0}(P) = g+1\qquad h_{1,1}(P)=4g+2.
\end{equation}
Here we see that (\ref{normal}) gives $N(P,T^6)=2g-4$ and
(\ref{newnormal}) gives $N(P,T^6)=2g+2$. The difference is $6$ and
these are clearly the translational modes along $T^6$ which must
exist for a generic cycle which breaks all the translational
symmetries. In total we find
\begin{equation}
N_B^L = 6g+6, \qquad N_B^R =4g+8 \ ,\qquad N_F^L = 12, \qquad N_F^R
= 4g+8,
\end{equation}

In all these cases one finds that $N(P,T^6) = 2h_{2,0}(P)$ as
predicted by (\ref{newnormal}). Let us make a comment on the first
two cases where the cycles are not ample. In these cases the
four-cycles preserve some of the symmetries of the torus and hence
the total number of translational Goldstone modes (equal to $2$ or
$4$ respectively) is less than $2h_{1,0}({\cal K})=6$. Nevertheless
we still find that the total number of normal modes is
$2h_{2,0}(P)$.

\section{Counting Black Holes}

Following \cite{Maldacena:1997de} we can obtain black hole solutions
of four-dimensional extended supergravity by further compactifying
the remaining spatial direction of the wrapped M5-brane on $S^1$. A
static wrapped M5-brane will be a magnetic source for the
four-dimensional Abelian gauge fields that arise from Kaluza-Klein
reduction of the M-theory three-form. One may also consider electric
charges by including M2-branes. According to Beckenstein and Hawking
the entropy of a macroscopic black hole is given by one quarter if
its horizon area. For the solutions at hand one finds
\cite{Maldacena:1997de}
\begin{equation}
S_{BH} = 2\pi\sqrt{\frac{1}{6}|q|\chi(P)},
\end{equation}
where $\chi(P) = 2 - 4h_{1,0}(P)+ 2h_{2,0}(P) + h_{1,1}(P)$ is the
Euler number of $P$. Here $q$ is the momentum carried by the
M5-brane along $S^1$, shifted by a contribution that arises from the
electric charges \cite{Maldacena:1997de}.

Following the  work of \cite{Strominger:1996sh} one obtains the
microscopic black hole degeneracy by counting all the modes of the
low energy M5-brane theory which preserve the supersymmetries of the
vacuum.  For an effective theory with $(0,4)$ supersymmetry this
amounts to counting the number of left moving modes with the right
movers in their vacuum. From Cardy's theorem this is determined by
the left-moving central charge and we find
\begin{equation}\label{macroentropy}
S = {2\pi\sqrt{\frac{1}{6}|q|c_L}} =
2\pi\sqrt{\frac{1}{6}|q|(\chi(P)+6h_{1,0}(P))},
\end{equation}
where we have used (\ref{ourcount}). For a generic Calabi-Yau, where
$h_{1,0}(P) = 0$, we see that the entropy is precisely reproduced by
a microscopic counting of the degrees of freedom of the M5-brane,
including the electric charges which in the microscopic picture
arise from shifts of the vacuum energy \cite{Maldacena:1997de}.
However for ${\cal K}=K3\times T^2$ or ${\cal K} = T^6$ we find
$h_{1,0}(P) \ne 0$ and the two entropy calculations do not agree, as
was pointed out in \cite{LopesCardoso}.

There is a further discrepancy. It is possible to compute the left
and right central charges of the superconformal $(0,4)$ fixed point
of the M5-brane  using gravitational and R-symmetry anomalies
\cite{Harvey:1998bx,Kraus:2005vz}. These arguments give $c_L =
h_{2,0}(P)+h_{1,1}(P)+2-4h_{1,0}(P)$ and
$c_R=6(h_{2,0}(P)-h_{1,0}(P)+1)$. This correctly accounts for the
black hole entropy but also differs from our counting by
$6h_{1,0}(P)$ for both the left and right central charges.

We propose the following resolution. Our field content naturally
splits into that of a `pure' $(0,4)$ supersymmetric sector with
\begin{eqnarray}\label{ournewcount}
\nonumber {N}_B^L&=&  2h_{2,0}(P)+h_{1,1}(P)+2-4h_{1,0}(P)\\
\nonumber {N}_B^R &=&   4h_{2,0}(P)+4-4h_{1,0}(P)\\
  {N}_F^L &=&  0\\
  \nonumber {N}_F^R &=& 4h_{2,0}(P)+4-4h_{1,0}(P),
\end{eqnarray}
and $h_{1,0}(P)$  $(4,4)$ multiplets with
\begin{equation}
{N}_B^L=4\qquad{N}_B^R=4\qquad{N}_F^L=4\qquad{N}_F^R=4.
\end{equation}
Note that we are not assuming that there is a left-moving
supersymmetry which acts on the $(4,4)$ multiplets, we are just
using them as a counting device.  The correct black hole degeneracy
and central charges are readily obtained if we only count the modes
of the $(0,4)$ sector. Furthermore both the black hole entropy and
anomaly arguments only count the degrees of freedom that are
massless at the conformal fixed point. Since the extra states that
we find fall into non-chiral $(4,4)$ multiplets it is reasonable to
conjecture that they become massive and hence do not appear in
spectrum of the conformal fixed.

We have not been able to provide any additional arguments to support
this proposal. However this claim is essentially a consequence of
our counting along with the results of
\cite{Harvey:1998bx,Kraus:2005vz}. To state this another way we note
that the quantum anomaly arguments determine the central charges at
the conformal fixed point and, combining this with our counting
(which is just a classical counting at lowest order), we deduce that
$h_{1,0}(P)$ $(4,4)$ multiplets become massive at the IR fixed
point.

We note that $(4,4)$ supersymmetry implies that the potential must
arise as the length-squared of a tri-holomorphic Killing vector on
the moduli space \cite{AlvarezGaume:1983ab}. When $h_{1,0}(P)\ne 0$
$\cal K$ has $U(1)$ isometries and these will induce Killing vectors
on the moduli space. We expect that, as a consequence of the
geometrical action of R-symmetry, the moduli space Killing vectors
should be tri-holomorphic. Therefore they can in principle lead to
the required  potential.

Let us make some comments on the mechanism that would provide such a
mass. One could object that the M5-brane moduli cannot become
massive because the equations of motion only involve derivatives of
the fields. In particular, although in section 2 we only gave the
lowest order equations of motion, the full non-linear equations have
been worked out for a general embedding into supergravity  and
indeed these only involve derivatives of the fields. These equations
of motion were derived in \cite{Howe} using the superembedding
formalism applied to the two derivative approximation to M-theory,
\ie\ standard eleven-dimensional supergravity of
\cite{Cremmer:1978km}. Another approach to obtaining the M5-brane
equations of motion comes from an analysis of the Goldstone modes of
the supergravity solution \cite{Adawi:1998ta}. The M5-brane
three-form $H$ is identified with zero-modes arising from gauge
transformations of the bulk three-form $C$. Again one would expect
that, as Goldstone modes, the equations of motion of the M5-brane
fields would only involve derivatives, even if higher derivative
terms were added to eleven-dimensional supergravity.

However there is an important caveat. It is well-known that at
next-to-leading order the M-theory effective Lagrangian contains the
anomaly $C\wedge I_8$ term \cite{Duff:1995wd}. This leads to a
source for $C$ and hence the also three-form $H$ on the M5-brane
worldvolume. Furthermore it is precisely the $C\wedge I_8$
term
in the effective action which is needed for cancelation of
the anomalies and which ultimately leads to the correct prediction
of the central charges in \cite{Harvey:1998bx,Kraus:2005vz}. Thus
one might suspect that this term induces a mass for the extra
$(4,4)$ multiplets that we have found.

We finish this section by noting that, in the examples above with
 $P=\Sigma\times T^2$  and $(4,4)$ supersymmetry, this mechanism
removes all the massless modes. Thus the M5-brane will behave as
though it is wrapped on a rigid cycle even though the cycle has
moduli.

\section{Conclusion}

In this paper we have discussed in detail the low energy dynamics of
an M5-brane wrapped on a smooth but otherwise arbitrary, complex
four-cycle in $K3\times T^2$ or $T^6$. In particular we gave the
lowest order equations of motion and determined the spectrum of
massless modes. This required a careful treatment of the zero-modes
that arise from translations along the $S^1$ factors and leads to a
spectrum in a agreement with supersymmetry. Finally we discussed the
counting of black hole microstates obtained by further reduction to
four-dimensions on another $S^1$. The naive counting of massless
modes does not reproduce the correct entropy and is not in agreement
with anomaly cancelation arguments. To resolve this we proposed that
$h_{1,0}(P)$ $(4,4)$ multiplets become massive and are removed from
the low energy spectrum. It would be very interesting to study this
mass mechanism in greater detail and verify that it indeed at work
here. In particular it would be interesting to incorporate the
effect of the $C\wedge I_8$ term on the M5-brane equations of
motion.

\section*{ Acknowledgements}

I am grateful to B. de Wit, T. Mohaupt, G. Moore and S. Trivedi for
useful discussions. This work is supported in part by the PPARC
grant PP/C507145/1 and the EU grant MRTN-CT-2004-512194.

\end{document}